\documentclass[journal,10pt]{IEEEtran} 
\usepackage[figurename=Fig.]{caption}
\usepackage{graphicx} 
\usepackage{tikz}
\usepackage{titlesec}
\usepackage{float}
\usetikzlibrary{calc, positioning, shapes.geometric, decorations.pathreplacing}
\usepackage{array}
\usepackage{verbatim}
\usetikzlibrary{calc, shapes, arrows, positioning}
\usepackage{authblk}
\usepackage{blindtext}
\usepackage{filecontents} 
\usepackage{ifpdf}
\usepackage{graphicx}
\usepackage{tabulary}
\usepackage{amsmath,amsthm,amssymb}
\usepackage{amsmath}

\usepackage{kantlipsum}
\allowdisplaybreaks
\usepackage{cleveref}
\usepackage{lipsum}%
\usepackage{optidef}
\usepackage[english]{babel} 
\theoremstyle{plain}

\crefname{theorem}{Theorem}{theorem}
\crefname{lemma}{Lemma}{Lemmas}

\usepackage{cite}
\usepackage{dsfont}
\usepackage{dblfloatfix}
\usepackage[justification=centering]{caption}
\usepackage{color}
\usepackage{algorithm}
\usepackage{algorithmicx, algpseudocode}
\usepackage{algpseudocode}
\usepackage{etoolbox}
\usepackage{subcaption}
\usepackage{balance}
\usepackage{empheq,etoolbox}
\usepackage[utf8]{inputenc}
\usepackage{multirow}
\usepackage{float} 
\usepackage{amssymb}
\usepackage{pifont}
\floatstyle{plaintop}
\restylefloat{table}
\patchcmd{\subequations}
\usepackage{lipsum} 

\begin{document}
 \captionsetup[figure]{name={Fig.},labelsep=period}

\title{\fontsize{20}{12}\selectfont Mode Switching-based STAR-RIS with Discrete Phase Shifters}

\bstctlcite{IEEEexample:BSTcontrol}

\author{{MohammadHossein Alishahi, Ming Zeng, Paul Fortier, Ji Wang, Nian Xia and Gongpu Wang}
    \thanks{
    This work was supported in part by NSERC under Grants RGPIN-2021- 02636 and CRC-2022-00115, and in part by FRQNT under Grant 341270. (Corresponding author: Ming Zeng.)
    
    M. H. Alishahi, M. Zeng, and P. Fortier are with the Department of Electric and Computer Engineering, Laval University, Quebec City, QC, G1V 0A6, CA. email: mohammadhossein.alishahi.1@ulaval.ca; ming.zeng@gel.ulaval.ca; paul.fortier@gel.ulaval.ca.}

    \thanks{{Ji Wang is with the Department of Electronics and Information Engineering, Central China Normal University, Wuhan 430079, China. email: jiwang@ccnu.edu.cn.}} 

    \thanks{{Nian Xia is with the School of Computer and Electronic Information, Nanjing Normal University, Nanjing 210023, China. e-mail: nian.xia@nnu.edu.cn}} 
    
    \thanks{{Gongpu Wang is with the School of Computer and Information Technology, Beijing Jiaotong University, Beijing 100044, China. e-mail: gpwang@bjtu.edu.cn}}

    }

\maketitle
\begin{abstract}
The increasing demand for cost-effective, high-speed Internet of Things (IoT) applications in the coming sixth-generation (6G) networks has driven research toward maximizing spectral efficiency and simplifying hardware designs. In this context, we investigate the sum rate maximization problem for a mode-switching discrete-phase shifters simultaneously transmitting and reflecting reconfigurable intelligent surface (STAR-RIS)-aided multi-antenna access point network, emphasizing hardware efficiency and reduced cost. A mixed-integer nonlinear optimization framework is formulated for joint optimization of the active beamforming matrix, user power allocation, and STAR-RIS phase shift vectors, including binary transmission/reflection amplitudes and discrete phase shifters. To solve the formulated problem, we employ a block coordinate descent method, dividing it into three subproblems tackled using difference-of-concave programming and combinatorial optimization techniques. Numerical results validate the effectiveness of the proposed joint optimization approach, consistently achieving superior sum rate performance compared to partial optimization methods, thereby underscoring its potential for efficient and scalable 6G IoT systems. 
\end{abstract}

\IEEEpeerreviewmaketitle
\vspace{-0.35cm}
\section{Introduction}
\IEEEPARstart{T}{he} sixth-generation (6G) networks incorporate advanced technologies, including multi-antenna access points (APs) and beamforming, to address the growing demand and increasing data traffic driven by Internet of Things (IoT) applications, such as smart cities and healthcare \cite{10597395}. Despite these advancements, significant challenges persist, particularly in addressing coverage and security issues in complex communication environments with attenuated channels \cite{nassereddine2024applications}. To mitigate these limitations, reconfigurable intelligent surfaces (RIS) have emerged as a promising solution. RIS are arrays of passive reflecting elements capable of dynamically manipulating electromagnetic waves through discrete or continuous phase shifters, enabling enhanced control over the wireless propagation environment. Recent studies \cite{li2022sum, 9110889, alishahi2024energy, 10103541} have highlighted the potential of RIS to improve network performance. For example, Ming \textit{et al}. \cite{li2022sum} demonstrated substantial sum-rate improvements in uplink non-orthogonal multiple access networks by integrating continuous phase shift RIS. Nonetheless, the hardware complexity and cost associated with continuous phase shifts have driven interest in discrete phase shifter RIS \cite{9110889}, which offers a practical balance between system performance and implementation feasibility.

Regardless of being discrete or continuous, RIS is inherently limited to serving users located on their reflecting side. To overcome this limitation, simultaneous transmission and reflection of RIS (STAR-RIS) has been introduced, enabling coverage extension by transmitting and reflecting signals on both sides through three distinct protocols: energy splitting (ES), mode switching (MS), and time switching (TS) \cite{10747257, 9751144, qin2023joint, 10720523, 10702481}. For instance, the authors of \cite{10747257} maximize the sum rate for STAR-RIS assisted orthogonal frequency division multiplexing with continuous phase shifts, while \cite{9751144} focus on weighted sum rate maximization for ES STAR-RIS assisted networks with coupled phase shifters. Moreover, Qin \textit{et al}. \cite{qin2023joint} demonstrated that continuous phase shift STAR-RIS outperforms traditional RIS in maximizing total computational throughput in complex wireless environments, with the TS protocol proving the most effective, followed by ES and MS. Despite offering the lowest performance, the MS protocol has recently garnered interest due to its simplicity and minimal hardware requirements \cite{qin2023joint, 10673989}. For example, \cite{10673989} showed that the MS discrete phase shifts STAR-RIS can achieve a balanced outage probability and ergodic rate compared to the continuous phase shift counterpart, while significantly reducing the complexity and cost of implementation.

To the best of our knowledge, existing studies have not investigated the joint optimization of discrete amplitude and phase shifters for MS STAR-RIS in communication systems, particularly in the context of sum-rate maximization. This letter addresses this research gap by proposing a novel framework for maximizing the sum rate in MS discrete phase shift STAR-RIS aided multi-antenna AP systems, facilitating simplified deployment in 6G networks. The primary contributions are as follows:

\vspace{0.03cm}
\textbf{System Design}: We implement an MS discrete phase shift STAR-RIS aided multi-antenna AP network, representing a practical and efficient deployment scenario.

\vspace{0.03cm}
\textbf{Problem Formulation and Solution}: A mixed-integer nonlinear optimization problem is formulated to maximize the sum rate by jointly optimizing the users' power allocation, binary transmission and reflection amplitudes, discrete phase shifts of the STAR-RIS, and the active beamforming matrix. To solve this complex problem, block coordinate descent (BCD) is employed, decomposing it into three subproblems. These subproblems are addressed using difference-of-convex (DC) programming and combinatorial optimization techniques for optimizing the users' power, binary amplitudes, discrete phase shifts, and the active beamforming matrix, respectively.

\vspace{0.03cm}
\textbf{Numerical Validation}: Numerical results highlight the significant performance improvements achieved by jointly optimizing the users' power, STAR-RIS components, and active beamforming matrix, as compared to optimizing these resources individually.

\textbf{Notations:} Vectors are represented by bold lowercase letters, while bold uppercase letters denote matrices. The symbols \(\mathcal{O}\), \((\cdot)^H\), \((\cdot)^T\), \(|\cdot|\), \(\mathbb{C}^{a \times b}\), and \(\mathrm{diag}(\mathbf{a})\) correspond to computational complexity, Hermitian transpose, transpose, absolute value or cardinality, the set of complex matrices of dimension \(a \times b\), and a diagonal matrix constructed from the vector \(\mathbf{a}\), respectively.

\section{{System Model and Problem Formulation}}
\subsection{{System Model}}
\begin{figure}[htbp]
\vspace{-0.5cm}
\centering
  \includegraphics[keepaspectratio, width=0.5\textwidth]{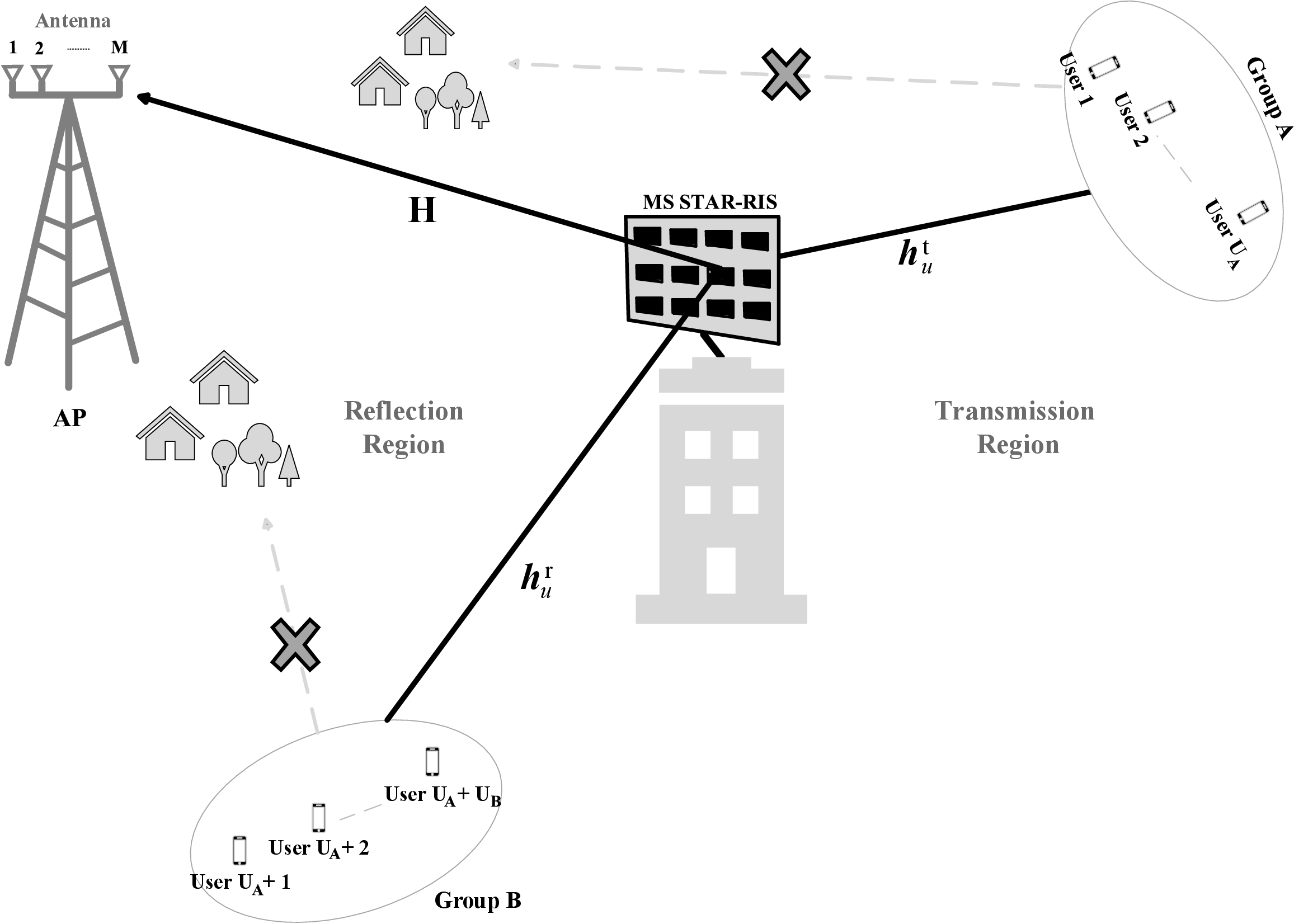}
  \caption{STAR-RIS aided uplink system.}
  \label{fig:1}
\end{figure}

Figure \ref{fig:1} illustrates an uplink STAR-RIS aided communication system, which includes a \( M \)-antenna AP, two groups of single-antenna users \( \mathcal{U}_\text{A} = \{1, \dots, U_\text{A}\} \) and \( \mathcal{U}_\text{B} = \{U_\text{A} + 1, \dots, U_\text{A} + U_\text{B}\} \), and a \( N \)-elements STAR-RIS indexed by \( \mathcal{N} = \{1, \dots, N\} \). In the absence of direct links due to severe blockages or harsh propagation environments, each user transmits the signal $s_u$ with power $p_u$ through STAR-RIS reflection or transmission regions, denoted by \( \mathcal{X} \in \{\text{t}, \text{r}\} \). The matrix \( \mathbf{H} \in \mathbb{C}^{N \times M} \) represents the channel coefficients between the STAR-RIS and AP, while \( \mathbf{h}^{\mathcal{X}}_{u} \in \mathbb{C}^{1 \times N} \) denotes the transmission or reflection channel gains from device \( u \in \mathcal{U}_\text{A}, \mathcal{U}_\text{B} \) to STAR-RIS. The active beamforming matrix, denoted as 
\(\mathbf{W} = [\mathbf{w}_{1}, \mathbf{w}_{2}, \dots, \mathbf{w}_{(U_\text{A} + U_\text{B})}] \in \mathbb{C}^{M \times (U_\text{A} + U_\text{B})}\), 
facilitates communication, where \(\mathbf{w}_{u} \in \mathbb{C}^{M \times 1}\) corresponds to each user \(u \in \mathcal{U}_\text{A}, \mathcal{U}_\text{B}\). Since users are located exclusively in either the transmission or reflection region, the channels 
\(\mathbf{h}^\text{t}_{u}\) for all \(u \in \mathcal{U}_\text{B}\) and 
\(\mathbf{h}^\text{r}_{u}\) for all \(u \in \mathcal{U}_\text{A}\) 
are set to zero.

This work considers discrete phase shifters and the MS protocol for STAR-RIS due to their ability to simplify hardware design and enhance practicality through straightforward phase control with predefined quantization levels, making it an ideal practical scenario \cite{10740607}. Given $\alpha^{\mathcal{X}}_{ n}, ~\forall n \in \mathcal{N}$, representing the amplitude coefficients, and $\theta^{\mathcal{X}}_{ n}$, denoting the corresponding discrete phase shifts in MS STAR-RIS, the phase shift vectors for both transmission and reflection sides can be expressed as\:
\begin{equation}
    \textbf{$\boldsymbol \phi$}^{\mathcal{X} }  = \bigg{(}\sqrt{\alpha^{\mathcal{X} }_{ 1}}e^{j\theta^{\mathcal{X} }_{ 1}},...,\sqrt{\alpha^{\mathcal{X} }_{ N}}e^{j\theta^{\mathcal{X} }_{ N}}\bigg{)}^T,
\end{equation}
where $\alpha^{\text{t}}_{n} + \alpha^{ \text{r}}_{n} = 1$, $\alpha^{\mathcal{X} }_{ n} \in \{0,1\}$, $\theta^{\mathcal{X}}_{n} \in  \{0, \frac{2 \pi}{Q}, ..., \frac{2 \pi (Q -1)}{Q} \}, \forall n \in \mathcal{N}$, $\mathcal{X} \in \{\text{t}, \text{r}\}$, and $Q$ denoting the quantization level \cite{10673989}.

To ensure coverage for both groups and prevent the extreme case where one region is deprived of sufficient resources, the following constraint imposes a minimum allocation of RIS elements to each region. This design choice is motivated by practical deployment considerations to maintain balanced quality of service across the network.
\begin{equation}
    \sum_{n = 1}^{N} \alpha^{\mathcal{X}}_n \geq \frac{N}{3}, \mathcal{X} \in \{\text{t}, \text{r}\}.
\end{equation}

The detected signal for the \(u\)-th user after applying the receive beamforming vector \(\mathbf{w}_u \) at the AP can be written as
\begin{align}
    y_u &= \mathbf{y} \mathbf{w}_u= \left( \sum_{i=1}^{U_\text{A} + U_\text{B}} \mathbf{h}_i \sqrt{p_i} s_i + \mathbf{N}_0 \right)\mathbf{w}_u \nonumber \\
        &= \mathbf{h}_u \mathbf{w}_u \sqrt{p_u} s_u + \sum_{i \neq u} \mathbf{h}_i \mathbf{w}_i \sqrt{p_i} s_i + \mathbf{n}_0 \mathbf{w}_u,
\end{align}
where \(\mathbf{h}_i = \sum_{\mathcal{X} \in \{\mathrm{t}, \mathrm{r}\}} \mathbf{h}_i^{\mathcal{X}} \mathrm{diag}(\boldsymbol{\phi}^{\mathcal{X}}) \mathbf{H}\) denotes the uplink channel between the AP and user $i$, whereas the noise vector $\mathbf{n}_0 \sim \mathcal{CN}(\mathbf{0}, \sigma^2 \mathbf{I})$ represents the additive white Gaussian noise at the AP.

All users simultaneously transmit their data to the AP, where MS STAR-RIS elements serve both reflection and transmission regions, causing each user to face intra-cluster and inter-cluster interference, denoted as:
\begin{equation}
    \mathcal{J}_{u} = \displaystyle \sum_{{m \notin \mathcal{G}, m \not = u}} |\sqrt{p_{m}} \textbf{h}_m \textbf{w}_{u}|^2 + \sum_ {{m = \mathcal{G}}} |\sqrt{p_{m}} \textbf{h}_m \textbf{w}_{u}|^2,
\end{equation}
where 
\begin{equation}
\mathcal{G} =
  \begin{cases}
  \mathcal{U}_{\text{B}} ~~ \text{if} & u \in \mathcal{U}_{\text{A}}, \\
  \mathcal{U}_{\text{A}} ~~ \text{if} & u \in \mathcal{U}_{\text{B}}. \\
  \end{cases}
\end{equation}
Therefore, the achievable sum rate is given by
\begin{equation}
    R_{\text{sum}} = \sum_{u = 1}^{U_\text{A} + U_\text{B}} \log_2 \bigg{(}1+\frac{| \sqrt{p_{u}}\textbf{h}_u \textbf{w}_{u}|^2}{\mathcal{J}_{u} + |{\sigma}\mathbf{w}_{u}|^2}\bigg{)},
\end{equation}

\subsection{Problem Formulation}
We formulate the following mixed integer sum rate maximization problem for a multi-antenna AP uplink network enhanced by an MS discrete phase shifts STAR-RIS.
\begin{subequations} \label{main_problem}
\begin{align}
\max_{p_u, {\boldsymbol \phi}^{\mathcal{X}}, \mathbf{W}} \quad & R_{\text{sum}} \\&\ \hspace{-1.6cm} \text{s.t.}~\textbf{$\alpha$}^{ \text{t}}_{n} + \textbf{$\alpha$}^{ \text{r}}_{n} = 1, ~\forall n \in \mathcal{N}, \\
& \hspace{-1cm} \textbf{$\alpha$}^{\text{t}}_{n}, \textbf{$\alpha$}^{ \text{r}}_{n} \in \{0, 1\},~\forall n \in \mathcal{N}, \\
& \hspace{-1cm} \sum_{n = 1}^{N} \alpha^{\text{t}}_n \geq \frac{N}{3},~\sum_{n = 1}^{N} \alpha^{\text{r}}_n \geq \frac{N}{3},\\
& \hspace{-1cm} \theta^{\mathcal{X}}_{n} \in  \{ 0, \frac{2 \pi}{Q}, ..., \frac{2 \pi (Q -1)}{Q} \}, \forall n \in \mathcal{N}, \mathcal{X} \in \{\text{t}, \text{r}\},\\
& \hspace{-1cm} ||\textbf{W} ||_F^2 \leq 1, \\
& \hspace{-1cm} 0 < p_{u} \leq p^{\text{max}}_{u}, ~\forall u \in \mathcal{U}_\text{A}, \mathcal{U}_\text{B},
\end{align}
\end{subequations}
where (\ref{main_problem}\rm{b} - \ref{main_problem}\rm{d}) constrain the amplitudes of the STAR-RIS elements in MS mode for both transmission and reflection regions; (\ref{main_problem}\rm{e}) regulates the discrete phase shifts; (\ref{main_problem}\rm{f}) restricts the gain of active beamformers at the AP, whereas (\ref{main_problem}\rm{g}) imposes a limit on the maximum power allocated to each user, denoted by $p^{\text{max}}_{u}$. Note that positive power allocation for each user, along with constraint (\ref{main_problem}d), ensures data rate fairness, preventing any user from being deprived of communication.

\section{Proposed Solution}
The non-convexity of the mixed integer optimization problem is evident due to the non-convexity of the objective function and constraints (\ref{main_problem}\rm{c}) and (\ref{main_problem}\rm{e}). To address this, we iteratively update the power allocation, phase-shift vectors of the STAR-RIS, and active beamforming matrix through the following subproblems. A concise flow chart of the proposed method is provided in Figure 2.

\textbf{Update $p_u \forall u$}: We simplify the optimization problem to determine the optimized power allocation for each user under the given phase shift vectors of the STAR-RIS and beamforming matrix, as follows.
\begin{subequations} \label{P_1 p_u}
\vspace{-0.2cm}
\begin{align}
\max_{\textbf P} \quad & F(\textbf P) \\&\ \hspace{-0.1cm}\text{s.t.}~ 0 < p_{u} \leq p^{\text{max}}_{u}, ~\forall u \in \mathcal{U}_\text{A}, \mathcal{U}_\text{B},
\end{align}
\end{subequations}
where  $\textbf P = (p_1, p_2, ..., p_{U_\text{A} + U_\text{B}})$,
\begin{equation}\label{DC}
\resizebox{0.99\columnwidth}{!}{$
F(\textbf P) =
\underbrace{\sum_{u = 1}^{U_\text{A} + U_\text{B}}
\log_2 \!\big({\mathcal{J}^{\text{tot}}_{u} + |\sigma \mathbf{w}^*_{u}|^2}\big)}_{f_1(\textbf P)}
-
\underbrace{\sum_{u = 1}^{U_\text{A} + U_\text{B}}
\log_2 \!\big({\mathcal{J}_{u} + |\sigma \mathbf{w}^*_{u}|^2}\big)}_{f_2(\textbf P)},
$}
\end{equation}
and 
\begin{equation} \label{lam}
    {\mathcal{J}^{\text{tot}}_{u} } = \displaystyle \sum_{m = \mathcal{U}_{\text{A}},\mathcal{U}_{\text{B}}} |\sqrt{p_{m}} \textbf{h}_m \textbf{w}_{u}|^2,  ~\forall u \in \mathcal{U}_{\text{A}},  \mathcal{U}_{\text{B}}.
\end{equation} 
The objective function is the difference of two concave functions with respect to $\textbf{P}$, thereby DC programming can effectively solve the following reformulated optimization problem in the $\iota_1$-th iteration until the objective function of \eqref{DC} no longer improves \cite{10720523}. 
\begin{subequations} {\label{DC1}}
\begin{align}
\max_{\textbf P} &\  f_1(\textbf P) -  f_2(\textbf P^{(\iota_1-1)}) -  \ \langle \nabla f_2(\textbf P^{(\iota_1-1)}), \textbf P - \textbf P^{(\iota_1-1)} \; \rangle \ \\
 & \hspace{-.6cm}\ \text{s.t.} \ 0 < p_{u} \leq p^{\text{max}}_{u}, ~\forall u \in \mathcal{U}_\text{A}, \mathcal{U}_\text{B}.
\end{align}
\end{subequations}

To optimize the phase shift vectors of the STAR-RIS and the active beamforming matrix, we leverage the fractional polynomial method discussed in \cite{10757366}. Accordingly, the objective function in \eqref{main_problem} can be reformulated as follows:
\begin{flalign} \label{G1}
\begin{aligned}
& R_{\text{sum}} = \sum_{u = 1}^{U_\text{A} + U_\text{B}} \log_2  (1+\mu_u) - \mu_u  \\
& + 2(\sqrt{1+\mu_u}) \Re \{ \bar \lambda_u \sqrt{p^*_u} \textbf{h}_{u} \mathbf{w}_u\} -|\lambda_u|^2 ({ \mathcal{J}^{\text{tot}}_{u} } + |\sigma \mathbf{w}_u|^2),
\end{aligned}
\end{flalign}
where $\mu_u$ and $\lambda_u$ are auxiliary variables and can be easily obtained from 
\begin{equation} \label{mu}
     \mu_u = \frac{|\sqrt{p^*_{u}} \textbf{h}_u \textbf{w}_{u}|^2}{ {\mathcal{J}_{u}} + {|\sigma \mathbf{w}_u |^2}}, ~\forall u \in \mathcal{U}_{\text{A}},  \mathcal{U}_{\text{B}},
\end{equation}
\begin{equation} \label{lambda}
     \lambda_u = \frac{(\sqrt{1+\mu_u}) \sqrt{p^*_{u}} \textbf{h}_u \textbf{w}_{u}}{  \mathcal{J}^{\text{tot}}_{u}  + |\sigma \mathbf{w}_u|^2}, ~\forall u \in \mathcal{U}_{\text{A}},  \mathcal{U}_{\text{B}}.
\end{equation}
By obtaining the auxiliary variables, we can update the phase shift vectors of the STAR-RIS and the active beamforming matrix as follows.

\textbf{Update ${\boldsymbol \phi}^{\mathcal{X}}$}: The optimized $\phi^{\mathcal{X}}$ that maximizes \eqref{G1}, can be obtained from the equivalent expression below \cite{10757366}
\begin{equation}
    R_{\text{sum}}^{\phi} = {\displaystyle \sum_{\mathcal{X} \in \{\text{t}, \text{r}\}} 2\Re \{ \mathbf{\omega}_{\mathcal{X}} \phi^{\mathcal{X}}\}} - \textbf{$\phi^{\mathcal{X}}$}^H \Omega_{\mathcal{X}} ~\phi^{\mathcal{X}},
\end{equation}
where 
\begin{flalign} \label{A_Label}
\begin{split}
    \mathbf{\omega}_{\mathcal{X}} &= \sum_{u = 1}^{U_\text{A} + U_\text{B}} (\sqrt{1+\mu^*_u}) \sqrt {p^{*}_u} \bar \lambda_u \mathbf{h}^{\mathcal{X}}_{u}\text{diag}(\mathbf{H}\mathbf{w}^{*}_u)  \in \mathbb{C}^{1 \times N},
\end{split}
\end{flalign}
\vspace{-0.7cm}
\begin{flalign} \label{w-Omega}
\begin{aligned}
     \mathbf{\Omega}_{\mathcal{X}} &= \sum_{u = 1}^{U_\text{A} + U_\text{B}} |\lambda_u|^2 \times \\
     & \hspace{-1.1cm}  \displaystyle \sum_{i \in \mathcal{U}_{\text{A}},\mathcal{U}_{\text{A}}} p^{*}_i~\text{diag}(\mathbf{w}^{H*}_u \mathbf{H}^H) {\mathbf{h}^{\mathcal{X}^H}_i} \mathbf{h}^{\mathcal{X}}_{i} \text{diag}( \mathbf{H} \mathbf{w}^*_u) \in \mathbb{C}^{N \times N}.
\end{aligned}
\end{flalign}
Therefore,  we can transform \eqref{G1} into the following to optimize $\phi^{\mathcal{X}}$.
\begin{subequations} {\label{phi}}
\begin{align}
\max_{{\boldsymbol \phi}^{\mathcal{X}}} &\ R_{\text{sum}}^{\phi} \\
  &\ \hspace{-1cm} \text{s.t.}~(\ref{main_problem}\rm{b} - \ref{main_problem}\rm{e}).
\end{align}
\end{subequations}
To solve the above mixed-integer nonlinear problem, we first set the total derivative of \( R_{\text{sum}}^{\phi} \) with respect to \( {\boldsymbol \phi}^{\mathcal{X}}\) to zero, which simplifies to 
\[
\boldsymbol{\phi}^{\mathcal{X}^*} = \mathbf{\Omega}_\mathcal{X}^{-1} \mathbf{\omega}^H_\mathcal{X}.
\]
To enforce the discrete phase constraint in (\ref{phi}\rm{e}), each phase of this stationary point, i.e., \( \theta_n^{\mathcal{X}} \), is projected to the nearest discrete value from the following set: 
\begin{equation} \label{theta}
  \theta_n^{\mathcal{X}^*} = \arg\min_{\theta \in \left\{ 0, \frac{2\pi}{Q}, \dots, \frac{2\pi(Q-1)}{Q} \right\}} \lvert \arg(\boldsymbol{\phi}^{\mathcal{X}^*}_n) - \theta \rvert.  
\end{equation}
Once (\ref{phi}\rm{e}) is relaxed, the following mixed-integer linear optimization problem can obtain the optimized transmission and reflection amplitudes, using MOSEK in MATLAB. 
\begin{subequations} {\label{phi1}}
\begin{align}
\max_{\boldsymbol{\alpha}^{\mathcal{X}}} &\ R_{\text{sum}}^{\phi} \\
  &\ \hspace{-1cm} \text{s.t.}~(\ref{main_problem}\rm{b} - \ref{main_problem}\rm{d}).
\end{align}
\end{subequations}

\textbf{Update $\mathbf{W}$}: Given power allocation to each user and phase shift vectors of STAR-RIS, we can easily obtain the optimized active beamforming matrix from the convex optimization problem below, using CVX in MATLAB.
\begin{subequations} {\label{w}}
\begin{align}
\max_{\textbf{W}} &\ 
\begin{aligned}[t]
 \sum_{u = 1}^{U_\text{A} + U_\text{B}} 2(\sqrt{1+\mu^*_u}) \Re \{\bar \lambda^*_u \sqrt{p^*_u} \textbf{h}^*_{u} \mathbf{w}_u\} \\
\hspace{-1cm} - |\lambda^*_u|^2 ({ \mathcal{J}^{\text{tot}}_{u} } + |\sigma \mathbf{w}_u|^2)
\end{aligned} \\
\text{s.t.} &\ ||{\textbf{W}}||_{\text{F}}^2\leq 1,
\end{align}
\end{subequations}

The following pseudo-code concisely summarizes the proposed method for obtaining the optimized solution.
\begin{figure}[t!]
\centering
\begin{tikzpicture}[thick, scale=0.25, every node/.style={scale=0.75}]
    \node[ellipse, draw, shape aspect=4, text width=10em, minimum height=4em] 
    at (-17.9, 14.6) (block1) {\begin{tabular}{c} \textbf{Problem Formulation} \\ $(p_{u},  \textbf{$\boldsymbol \phi$}^{\mathcal{X}}, \textbf{W})$ \end{tabular}};
    
    \node[ellipse, draw, shape aspect=4, text width=3.3em, minimum height=4em] 
    at (5, 14.6) (block2) {\begin{tabular}{c} \textbf{$R_{\text{sum}}$} \end{tabular}};
frase    
    \node[ellipse, draw, black, dotted, shape aspect=4, text width=10em, minimum height=3em, align=center] 
    at (-17.2, 4.8) (block3) {\begin{tabular}{c} \textbf{Optimize $p_u$} \\ \textbf{via DC Programming} \end{tabular}};
    
    \node[ellipse, draw, black, dotted, shape aspect=4, text width=10em, minimum height=3em, align=center] 
    at (1.2, 4.8) (block4) {\begin{tabular}{c} \textbf{Optimize $ \textbf{$\boldsymbol \phi$}^{\mathcal{X} }$} \\ \textbf{via MOSEK} \end{tabular}};
    
    \node[ellipse, draw, black, dotted, shape aspect=4, text width=10em, minimum height=3em, align=center] 
    at (-8, 9.4) (block5) {\begin{tabular}{c} \textbf{Optimize $\textbf{W}$} \\ \textbf{via CVX} \end{tabular}};

    \node[ellipse, draw, black, dotted, shape aspect=4, text width=10em, minimum height=3em, align=center] 
    at (-8, 0.4) (block51) {\begin{tabular}{c} \textbf{Optimize ${\mu_u}$ and  ${\lambda_u}$} \\ \textbf{via Closed Form} \end{tabular}};
    
    \node[ellipse, draw, dotted, red, minimum width=33em, minimum height=13em] 
    at (-8, 5) (block6) {};
    
    \draw[thick, line width=1.5pt, gray, ->] 
        (block1.south west) to[bend right=20] (block6.west);
    \draw[thick, line width=1.5pt, red!50!, ->] 
        (block3.south) to[bend right=30] (block51.west);
    \draw[thick, line width=1.5pt, red!50!, ->] 
        (block4.north) to[bend right=30] (block5.east);
    \draw[thick, line width=1.5pt, red!50!, ->] 
        (block5.west) to[bend right=30] (block3.north);
    \draw[thick, line width=1.5pt, gray, ->] 
        (block6.east) to[bend right=20] (block2.south east);

        \draw[thick, line width=1.5pt, red!50!, ->] 
        (block51.east) to[bend right=30] (block4.south);
\end{tikzpicture}
\caption{Flowchart of the proposed optimization method.}
\label{fig2}
\end{figure}
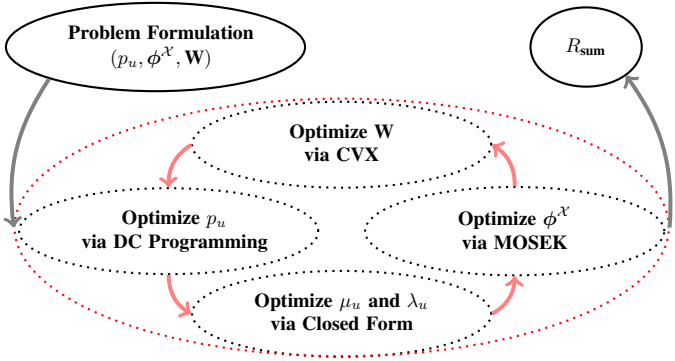
\begin{algorithm} [htbp]
\vspace{-0.15cm}
\setlength{\abovecaptionskip}{0pt}
\setlength{\belowcaptionskip}{0pt}
\caption{ Proposed Iterative Framework }
\label{I}
\begin{algorithmic}[1]
     \State Initialize $\textbf{P}$, $\textbf{$\boldsymbol \phi$}^{\mathcal{X}}$,  $\textbf{W}$, and $R^{1}_{\text{sum}} = 0$.
     \Repeat
     \State Update  $\textbf{P}$ from \eqref{DC1}.
     \State Calculate  $\mu_u~\forall u$ and $\lambda_u~\forall u$ from \eqref{mu} and  \eqref{lambda}.
     \State Update  $\textbf{$\boldsymbol \phi$}^{\mathcal{X}}$ from \eqref{theta} and \eqref{phi1}.
     \State Obtain  $\textbf{W}$ from \eqref{w}.
     \State Calculate $R^{\iota_2 + 1}_{\text{sum}}$.
     \State $\iota_2 \leftarrow \iota_2 + 1$,
     \Until{{$R^{\iota_2 + 1}_{\text{sum}} - R^{\iota_2}_{\text{sum}}> \epsilon$.}}

\end{algorithmic}
\end{algorithm}

\subsection{Convergence and Computational Complexity}
\begin{figure}[htbp]
\vspace{-0.2cm}
    \centering
    \includegraphics[width=0.7\linewidth]{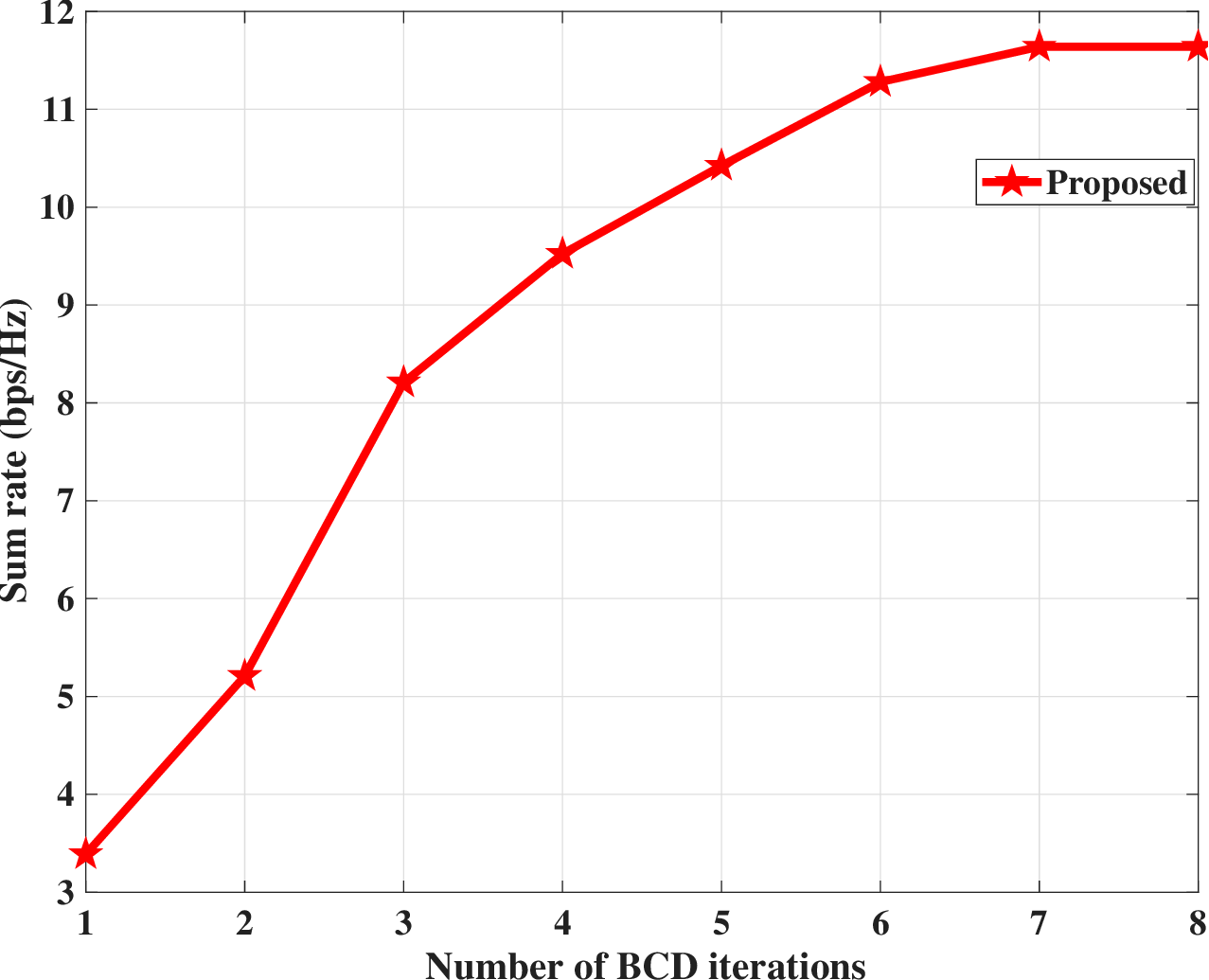}
    \caption{{Sum rate vs. iteration index.}}
    \label{fig:31}
    \end{figure}

The proposed iterative framework combines DC programming and combinatorial optimization techniques. The optimization subproblems in \eqref{DC1} and \eqref{w} converge, because of the convexity of these subproblems \cite{10720523}, while the mixed-integer linear problem ensures convergence owing to its finite feasible region \cite{wolsey2020integer}. Overall, the entire algorithm converges as it terminates once the main objective function no longer improves. To illustrate practical convergence, Fig.~\ref{fig:31} plots the sum rate versus iteration index, showing that the proposed BCD-based algorithm converges rapidly within a few iterations.
    
The computational complexity of the proposed algorithm consists of three main components: $\mathcal{O}(\gamma^2 \iota_1)$ for the DC programming stage, where $\gamma$ is the number of dual variables and $\iota_1$ is the number of DC iterations \cite{10720523}; $\mathcal{O}(\iota \cdot N^\beta)$ for solving the mixed-integer linear problem in \eqref{phi1}, where $\iota$ is the number of nodes explored in the branch-and-bound tree and $\beta \in [1.5, 3]$ typically depends on the problem structure, due to MOSEK’s use of branch-and-bound, cutting planes, and presolve heuristics \cite{wolsey2020integer}; and $\mathcal{O}\left((2(U_{\text{A}} + U_{\text{B}}) + 1)^{3.5} \log_2 \left(\frac{1}{\epsilon}\right)\right)$ for solving the convex optimization problem in \eqref{w}.
\begin{figure*}[t!]
\vspace{-0.4cm}
    \centering
    \begin{subfigure}[b]{0.32\textwidth}
        \includegraphics[width=\linewidth]{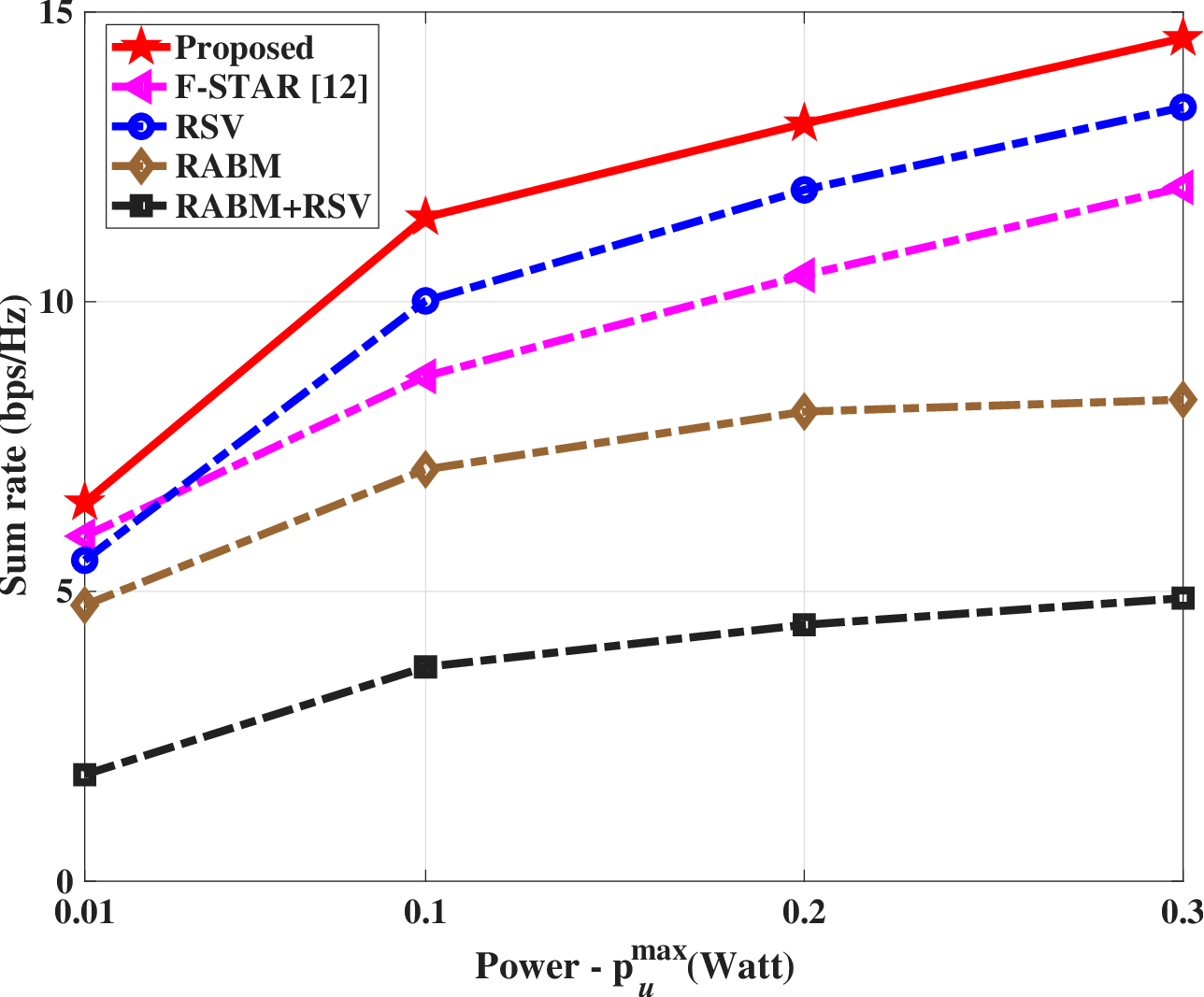}
        \caption{Sum rate vs. user maximum power.}
    \end{subfigure}
    \begin{subfigure}[b]{0.33\textwidth}
        \includegraphics[width=\linewidth]{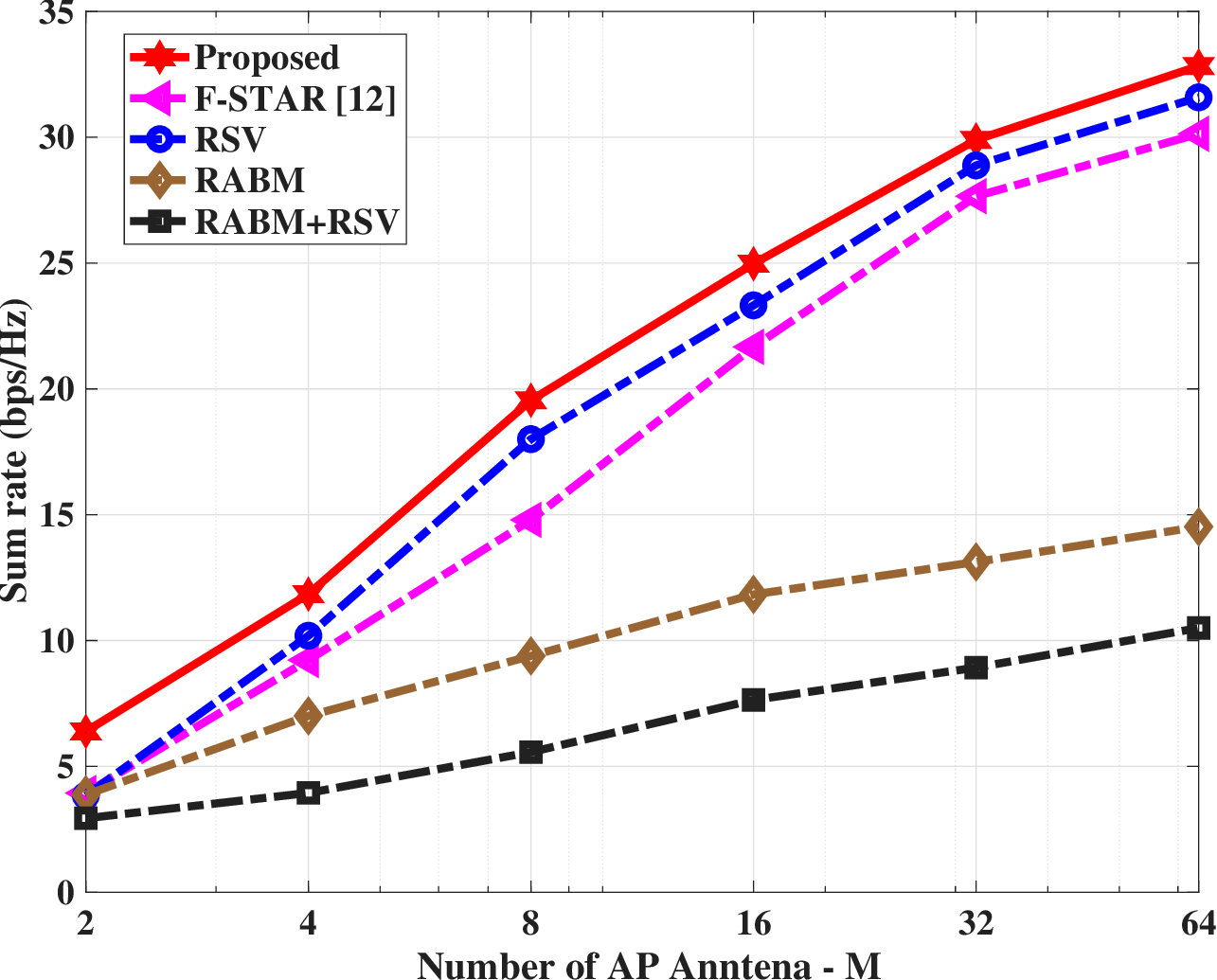}
        \caption{Sum rate vs. AP antennas.}
    \end{subfigure}
    \begin{subfigure}[b]{0.32\textwidth}
        \includegraphics[width=\linewidth]{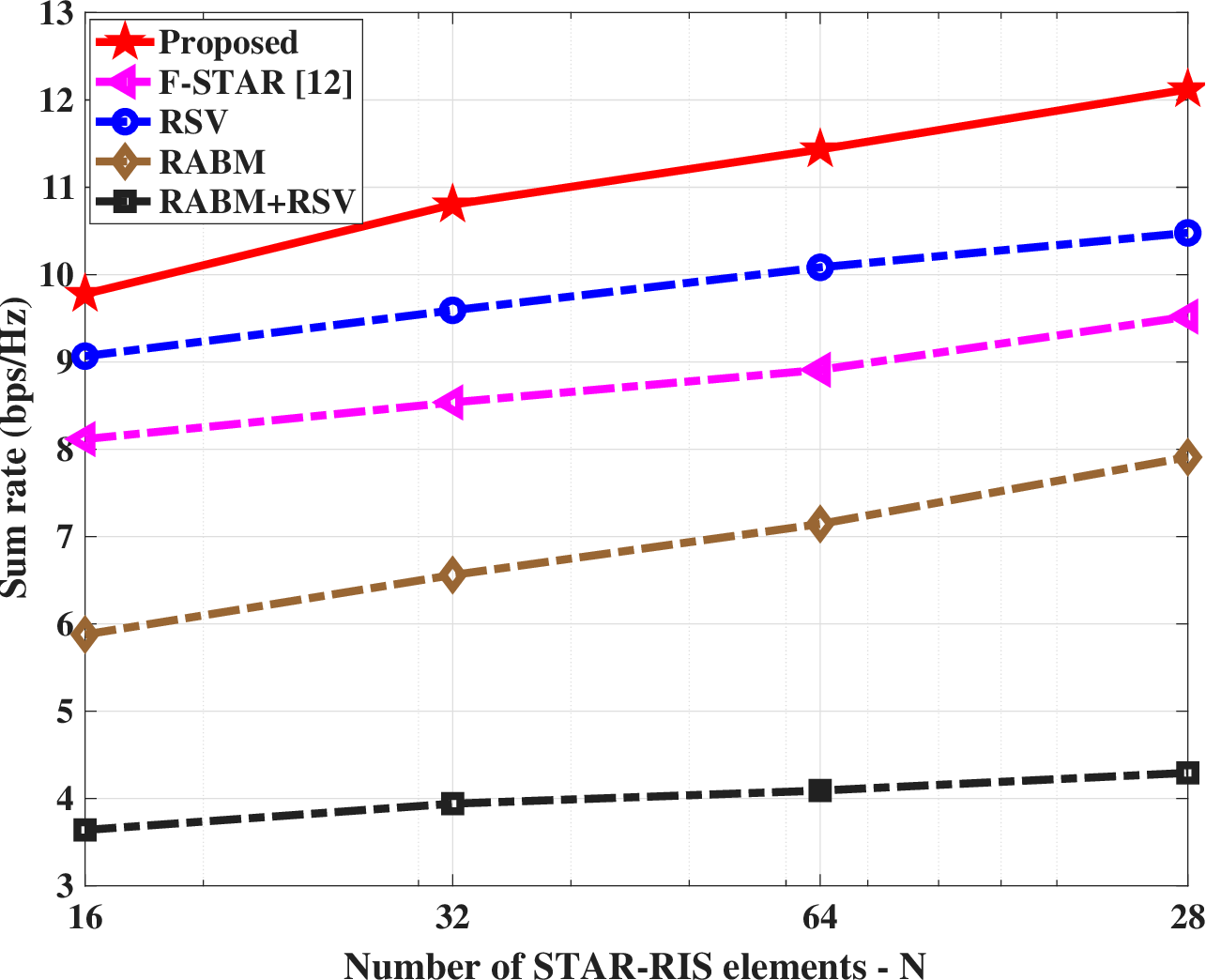}
        \caption{Sum rate vs. STAR-RIS elements.}
    \end{subfigure}
\caption{Sum rate versus different system parameters.}
     \label{fig4-5-6}
\end{figure*}

\vspace{-.2cm}
\section{Numerical Results}
For the default simulation parameters, we consider a $4$-antenna AP located at the origin in an uplink network aided by a $64$-element MS STAR-RIS positioned at $(75~\text{m}, 25 ~\text{m})$ with discrete phase shifts quantized in $8$ levels, communicating with two user groups, each consisting of $4$ users uniformly distributed along $20 ~\text{m}$ radius circles centered at $(100 ~\text{m}, 0 ~\text{m})$ and $(75 ~\text{m}, 50 ~\text{m})$, respectively. Note that equal user numbers are assumed for simplicity, while the algorithm handles unequal distributions since neither the formulation nor the solution imposes constraints. Each user has a maximum transmit power of $0.1$ W and experiences noise with a variance of $-100$ dBm.We consider a quasi-static flat-fading channel with large-scale path loss and small-scale Rayleigh fading, where the fading coefficients are i.i.d. complex Gaussian random variables with zero mean and unit variance. The path loss is modeled by $\rho d^{-\alpha}$, where $d$ is the distance, $\rho = -20$ dB, and $\alpha = 2.5$ is the path loss exponent between AP and STAR-RIS or STAR-RIS and users \cite{li2022sum, 10673989, 10044972}. The algorithm is run for $10^3$ iterations with an error tolerance of $\epsilon = 10^{-4}$. We compare the proposed method with four benchmarks: (1) joint optimization of users' power and phase shift vectors with a random active beamforming matrix (RABM), (2) joint optimization of users' power and beamforming matrix with random STAR-RIS vectors (RSV), (3) optimizing users' power with both random beamforming and STAR-RIS vectors (RABM+RSV), and (4) joint optimization of users' power, beamforming matrix, and STAR-RIS phase shift vectors with a fixed number of elements serving for transmission and reflection regions (F-STAR) \cite{10673989}. However, \cite{10673989} studied the probability of outage and ergodic rate for such a system.

Figure \ref{fig4-5-6}(a) illustrates an increasing trend in the sum rate performance for all schemes as \(p_{\text{max}}^u\) increases because users have more freedom to allocate additional power within their budget. A detailed comparison between RSV and RABM reveals that optimizing the active beamforming matrix has a more substantial impact than optimizing the STAR-RIS phase shift vectors, demonstrating the critical role of the number of AP antenna in enhancing sum rate performance. Furthermore, RSV outperforms F-STAR at medium- and high-power budgets due to the greater flexibility and adaptability of random STAR-RIS phase shift vectors to explore a wider range of amplitude combinations, while F-STAR offers symmetry and simplifies system design, highlighting a trade-off between sum rate performance and design simplicity. However, F-STAR with fixed-amplitude allocation at low-power budgets provides more stable and efficient performance, as RSV's flexibility becomes less effective under power-limited conditions. The proposed scheme consistently outperforms all benchmarks, highlighting the benefits of joint optimization of all resources. Its superior performance compared to RSV and F-STAR highlights the significance of jointly optimizing the transmission and reflection phase shifts and amplitudes of the STAR-RIS elements. Figure \ref{fig4-5-6}(b) shows the sum rate versus the number of AP antennas.  The widening gap between the schemes with an optimized beamforming matrix and those without highlights the significant impact of beamforming matrix optimization. In particular, F-STAR and RABM outperform RSV in single-antenna AP systems, emphasizing the effectiveness of STAR-RIS component optimization over beamforming matrix optimization in such configurations. Across all configurations, our proposed method remains superior. Figure \ref{fig4-5-6}(c) compares the performance of the sum rate for all schemes versus the increasing number of STAR-RIS elements, where our proposed method is still the best among all. This trend occurs because, as the number of STAR-RIS elements increases, the system can enhance signal quality by more effectively controlling the reflection and transmission of signals.

\begin{figure} [htbp]
\vspace{-0.1cm}
\centering
 \includegraphics[width=0.7\linewidth]{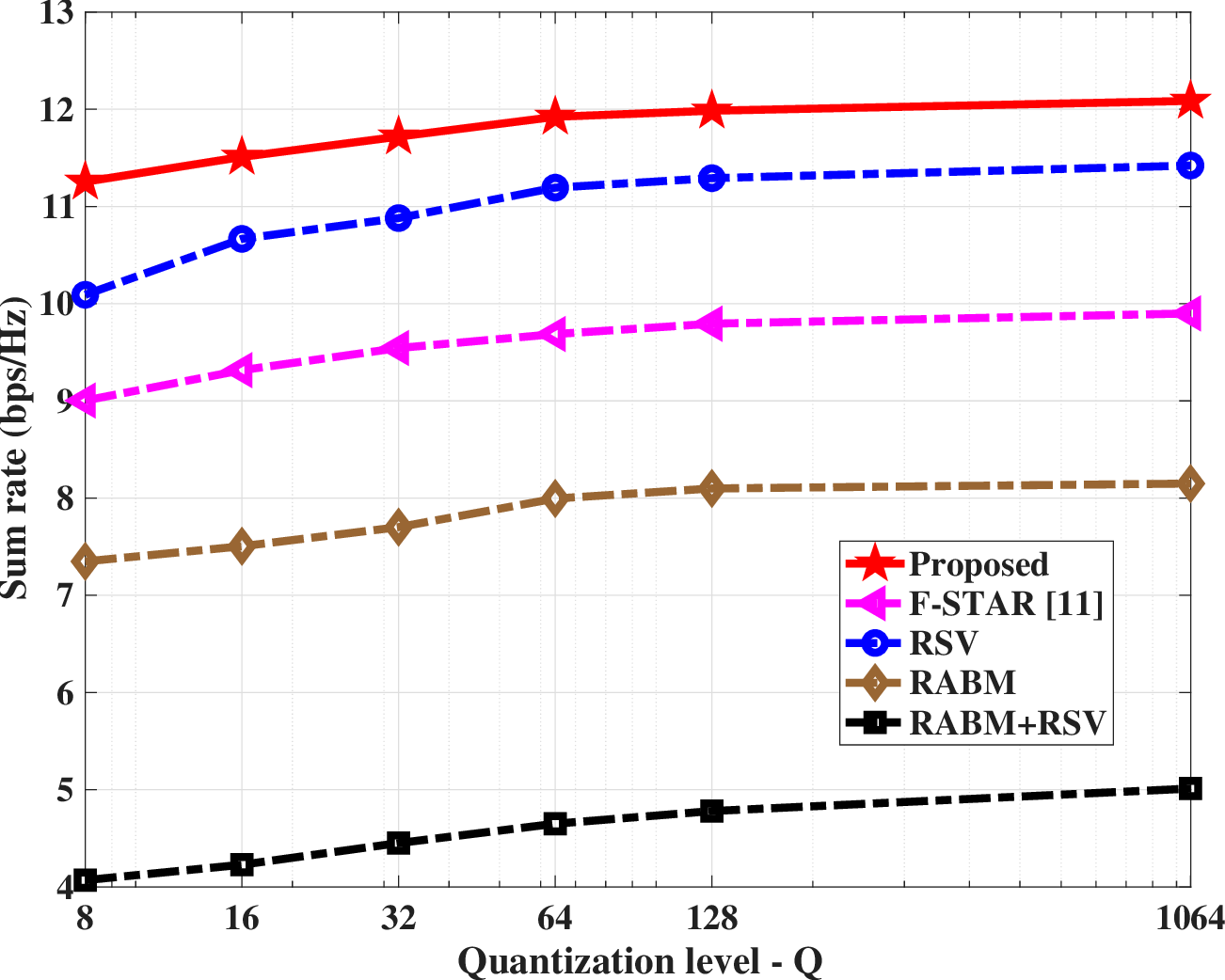}
    \caption{Sum rate vs. quantization level $Q$.}
  \label{fig7}
\end{figure}

Figure \ref{fig7} shows that the sum rate increases with the quantization level, approaching the continuous phase case (i.e., $Q \to \infty$). However, the gain becomes negligible at high quantization levels, indicating that moderate resolution provides efficient performance with simpler implementation. Our proposed scheme consistently outperforms all benchmarks for different $Q$.

\vspace{-0.5cm}
\section{Conclusion}
This letter investigated the impact of power allocation to each user, active beamforming matrix, and phase shift vectors of STAR-RIS optimization on the sum rate performance of MS discrete phase shift STAR-RIS-aided multi-antenna AP wireless systems. A mixed-integer nonlinear optimization problem was formulated to optimize these parameters jointly. The BCD method was used to solve three subproblems, using DC programming and combinatorial optimization techniques. Numerical results demonstrated the superiority of the proposed scheme over conventional benchmarks, validating the effectiveness of joint optimization. The adoption of MS discrete phase shift STAR-RIS enhances its practicality for 6G networks.

\vspace{-0.4cm}
\bibliographystyle{IEEEtran}
\bibliography{biblio}

\begin{thebibliography}{10}
\providecommand{\url}[1]{#1}
\csname url@samestyle\endcsname
\providecommand{\newblock}{\relax}
\providecommand{\bibinfo}[2]{#2}
\providecommand{\BIBentrySTDinterwordspacing}{\spaceskip=0pt\relax}
\providecommand{\BIBentryALTinterwordstretchfactor}{4}
\providecommand{\BIBentryALTinterwordspacing}{\spaceskip=\fontdimen2\font plus
\BIBentryALTinterwordstretchfactor\fontdimen3\font minus \fontdimen4\font\relax}
\providecommand{\BIBforeignlanguage}[2]{{%
\expandafter\ifx\csname l@#1\endcsname\relax
\typeout{** WARNING: IEEEtran.bst: No hyphenation pattern has been}%
\typeout{** loaded for the language `#1'. Using the pattern for}%
\typeout{** the default language instead.}%
\else
\language=\csname l@#1\endcsname
\fi
#2}}
\providecommand{\BIBdecl}{\relax}
\BIBdecl

\bibitem{10597395}
M.~Le \emph{et~al.}, ``Applications of distributed machine learning for the internet-of-things: A comprehensive survey,'' \emph{IEEE Communications Surveys and Tutorials}, pp. 1--1, Jul. 2024.

\bibitem{nassereddine2024applications}
M.~Nassereddine \emph{et~al.}, ``Applications of {I}nternet of {T}hings ({IoT}) in smart cities,'' pp. 109--136, Jan. 2024.

\bibitem{li2022sum}
X.~Li \emph{et~al.}, ``Sum rate maximization for {RIS}-aided {NOMA} with direct links,'' \emph{IEEE Networking Letters}, vol.~4, no.~2, pp. 55--58, Mar. 2022.

\bibitem{9110889}
B.~Di \emph{et~al.}, ``Hybrid beamforming for reconfigurable intelligent surface based multi-user communications: Achievable rates with limited discrete phase shifts,'' \emph{IEEE Journal on Selected Areas in Communications}, vol.~38, no.~8, pp. 1809--1822, Aug. 2020.

\bibitem{alishahi2024energy}
M.~Alishahi \emph{et~al.}, ``\textcolor{black}{Energy Minimization for {IRS}-Aided Wireless Powered Federated Learning Networks with {NOMA}},'' \emph{\textcolor{black}{IEEE Internet of Things Journal}}, vol.~11, no.~9, pp. 16\,339--16\,350, Jan. 2024.

\bibitem{10103541}
K.~Singh \emph{et~al.}, ``Performance analysis of {RIS}-assisted full-duplex communications with infinite and finite blocklength codes,'' \emph{IEEE Transactions on Communications}, vol.~71, no.~7, pp. 4262--4282, Jul. 2023.

\bibitem{10747257}
J.~Moon \emph{et~al.}, ``Sum rate maximization via {STAR}-{RIS} element allocation in {OFDM} systems,'' \emph{IEEE Transactions on Vehicular Technology}, pp. 1--6, Nov. 2024.

\bibitem{9751144}
H.~Niu \emph{et~al.}, ``Weighted sum-rate maximization for {STAR}-{RIS}s-aided networks with coupled phase-shifters,'' \emph{IEEE Systems Journal}, vol.~17, no.~1, pp. 1083--1086, Mar. 2023.

\bibitem{qin2023joint}
X.~Qin \emph{et~al.}, ``Joint resource allocation and configuration design for {STAR}-{RIS}-enhanced wireless-powered {MEC},'' \emph{IEEE Transactions on Communications}, vol.~71, no.~4, pp. 2381--2395, Jan. 2023.

\bibitem{10720523}
M.~Alishahi \emph{et~al.}, ``Total computational bits maximization for {STAR}-{RIS} aided wireless power transfer mobile edge computing networks: {TDMA} or {NOMA}?'' \emph{IEEE Transactions on Vehicular Technology}, vol.~74, no.~2, pp. 3345--3358, Feb. 2025.

\bibitem{10702481}
Y.~Li \emph{et~al.}, ``Weighted sum power maximization for {STAR}-{RIS} assisted {SWIPT} systems,'' \emph{IEEE Transactions on Wireless Communications}, vol.~23, no.~12, pp. 18\,394--18\,408, Dec. 2024.

\bibitem{10673989}
Y.~Liu \emph{et~al.}, ``{STAR}-{RIS} assisted wireless communications with discrete phase shifter,'' \emph{IEEE Wireless Communications Letters}, vol.~13, no.~11, pp. 3152--3156, Nov. 2024.

\bibitem{10740607}
H.~Li \emph{et~al.}, ``{STAR}-{RIS} in cognitive radio networks,'' \emph{IEEE Transactions on Wireless Comm.}, vol.~23, no.~12, pp. 19\,649--19\,666, Dec. 2024.

\bibitem{10757366}
M.~Alishahi \emph{et~al.}, ``Latency minimization for {STAR}-{RIS} aided federated learning networks with wireless power transfer,'' \emph{IEEE Internet of Things Journal}, vol.~12, no.~7, pp. 8508--8522, Apr. 2025.

\bibitem{wolsey2020integer}
L.~A. Wolsey, \emph{Integer programming}.\hskip 1em plus 0.5em minus 0.4em\relax John Wiley \& Sons, Sep. 2020.

\bibitem{10044972}
M.~{Alishahi} \emph{et~al.}, ``Energy minimization for wireless-powered federated learning network with {NOMA},'' \emph{IEEE Wireless Communications Letters}, vol.~12, no.~5, pp. 833--837, Feb. 2023.

\end{thebibliography}

\end{document}